\def\year{2020}\relax
\documentclass[letterpaper]{article} 
\usepackage{aaai20}  
\usepackage{times}  
\usepackage{helvet} 
\usepackage{courier}  
\usepackage[hyphens]{url}  
\usepackage{graphicx} 
\usepackage{graphicx}  
\usepackage{listings}
\usepackage{PDDL}
\nocopyright
\title{HTN Planning Domain for Deployment of Cloud Applications}
\author{Ilche Georgievski\\ 
Service Computing Department\\ 
Institute for Architecture of Application Systems\\
University of Stuttgart\\
firstname.lastname@iaas.uni-stuttgart.de 
}
\begin{document}

\maketitle

\begin{abstract}
Cloud providers are facing a complex problem in configuring software applications ready for deployment on their infrastructures. Hierarchical Task Network (HTN) planning can provide effective means to solve such deployment problems. We present an HTN planning domain that models deployment problems as found in realistic Cloud environments.
\end{abstract}

\section{Introduction}
The choice of enterprises to have their software applications deployed and run on Cloud infrastructures is motivated by elasticity, flexibility, scalability and high availability, which are the promised benefits of Cloud Computing~\cite{buyya2009:cloud-computing}. Before being able to deploy an application, Cloud management teams have to find, choose and configure appropriate software components that will compose the application, thus making it ready for deployment. Solutions to such deployment problems are typically configured manually or using predefined scripts. Both approaches seem impractical for Cloud management as they are error-prone and require strenuous effort to handle a large number of components, versions of components and high interdependence between components~\cite{binz2014:tosca-cloud-deployment}. Cloud providers and Cloud Computing community are therefore in need for approaches and tools that can solve deployment problems automatically~\cite{arshad2003:planningdeployment}. 


Artificial Intelligence (AI) planning can provide powerful means to automatically and efficiently search for solutions to deployment problems. Hierarchical Task Network (HTN) planning appears to be particularly suitable as it can incorporate the configuration knowledge otherwise provided by the Cloud management teams. In this paper, we describe HTN planning problems that correspond to deployment problems. We first introduce the component model used for describing deployment problems, and then we describe an HTN planning domain that models such deployment problems. 


\section{Component Model}
Aelous is a component model used to describe software applications as found in realistic Cloud deployments~\cite{dicosmo2014:aeolus}. A central element of Aeolus is a \emph{component}, a manageable software resource that provides and requires functionalities. Each component has three states: \emph{uninstalled}, \emph{installed}, and \emph{running}. State transitions are done using \emph{deployment actions}. For example, we can run an installed component by invoking a \emph{runComponent} action on it. A component may require or provide different functionality at each state. The requirement of functionality is expressed via \emph{require ports}, and providing of functionality through \emph{provide ports}. A component can transition from one state to another only if the functionality the new state requires can be provided by other component(s). When a component goes to a new state, its require ports are bound to appropriate provide ports of other component(s). This process is called \emph{port binding}. Once a component is in the new state, its provide ports become active via \emph{port activation}. Since a component transitions from some state, its ports must be deactivated and unbound via \emph{port deactivation} and \emph{port unbinding}. 

This component model has two interesting features. First, since component represent abstract entities, they must be instantiated. The peculiarity comes from the fact that the \emph{creation of new uninitialised component instances} happens on demand during runtime. The second feature is that a cycle may occur between states of different component instances: an instance is expected to provide a functionality, but it is not possible because the instance is required to change its state at the same time~\cite{lascu2013:planningdeployment}. We can deal with such cycles by creating as many instances of the same component as needed, and deploying them in different states. This process is called \emph{instance duplication}.

A \emph{configuration} describes all available components, currently deployed component instances and their states, and current bindings of components via ports. A \emph{deployment problem} consists of an initial configuration, a set of deployment actions, and a request for a new configuration, i.e., application. The solution is a \emph{deployment run}, which is a sequence of deployment actions on components that, when deployed, produce the required configuration.

\section{Deployment Planning Domain}
We now describe the \begin{footnotesize}\verb|deployment|\end{footnotesize} HTN domain model that encodes deployment problems. Our description is based on the paper in which we introduced the HTN planning approach to solving deployment~\cite{georgievski2017:cloudapps}.

\subsection{Components, States and Ports}
We describe components, instances, and ports using \begin{footnotesize}\verb|component instance port|\end{footnotesize} as domain types. Each component type would be represented as an object of \begin{footnotesize}\verb|component|\end{footnotesize}. For example, a {\em Wordpress} component would be represented as \begin{footnotesize}\verb|wordpress - component|\end{footnotesize}.

Even though Aelous associates components with states, component instances are the ones to be in a specific state during planning. We encode a state of an instance in a predicate ``{\em (state instance)}'', where {\em state} is a string representing the instance's state, and {\em instance} is a variable representing the component instance. For example, \begin{footnotesize}\verb|(installed w1)|\end{footnotesize} represents a Wordpress instance $\mathit{w1}$ in an installed state.

We encode the association of states with ports in a predicate ``{\em (statePort component port)}'', where {\em statePort} is a string describing the type of port in a specific state, and {\em component} is a variable referring to an abstract component that requires or provides a port represented by the {\em port} variable. For example, if {\em Wordpress} requires the $\mathit{httpd}$ port in the installed state, we encode it as \begin{footnotesize}\verb|(installed-require wordpress httpd)|\end{footnotesize}. Note that such knowledge holds for all instances of the respective component. These predicates are static predicates. 

\subsection{Deployment Actions}
We encode all deployment actions as planning actions as follows. Action's parameters correspond either to a component instance variable or to variables of a port and two instances in the case of binding actions (see below). The preconditions and effects of each action capture the semantics of the respective deployment action. Listing~\ref{lst:action} shows the action that corresponds to the {\em startComponent} deployment action, which makes the state of an instance to become installed. It uses a conditional effect within a universal quantifier to activate all the ports associated with the installed state of the component which the current instance belongs to. The encoding of the actions for running, stopping and terminating component instances are similar. There are also binding actions responsible for low-level binding of ports -- require ports are bound to provide ports. They are represented by two planning actions. The \begin{footnotesize}\verb|bind|\end{footnotesize} action creates a binding between a provide port of some instance and a require port of another one, and the \begin{footnotesize}\verb|unbind|\end{footnotesize} action deletes an already established binding between two instances.

\begin{lstlisting}[
	float=t,
	caption={HTN action for starting a component instance.},
	label={lst:action},
	language=PDDL]
(:action start
  :parameters (?i - instance)
  :precondition (not (installed ?i))
  :effect (and 
    (installed ?i) 
    (forall (?p - port) (when 
      (and (installed-provide ?c ?p) 
           (type ?i ?c))
      (active ?p ?i))
    )
  )
)
\end{lstlisting}

The last action is for creating new uninitialised instances. The \begin{footnotesize}\verb|createInstance|\end{footnotesize} action shown in Figure~\ref{lst:action2} uses a domain function to get (and increase) a number that we use to uniquely represent an instance in a predicate as \begin{footnotesize}\verb|(instance ?iNum - number)|\end{footnotesize}. The domain function does not take arguments and serves as a counter to keep track of the current value that can be assigned for new instances. The action uses another predicate, \begin{footnotesize}\verb|(type ?iNum - number ?c - component)|\end{footnotesize}, to associate a new instance with a particular component.

\begin{lstlisting}[
	float=t,
	caption={HTN action for creating an uninitialised component instance.},
	label={lst:action2},
	language=PDDL]
(:action createInstance
  :parameters (?c - component)
  :precondition ()
  :effect (and 
     (instance (instance-number))
     (type (instance-number) ?c)
     (increase (instance-number) 1)
  )
)
\end{lstlisting}

\subsection{Configuration Processes}
We now describe the encoding of processes needed for configuring applications. The basic process requires satisfaction of dependencies to functionalities provided by components. Let us assume that an instance in an uninstalled state cannot have requirements to be satisfied. We may then consider two abstractions of the basic process. The first one refers to acquiring a component functionality in an installed state, while the second abstraction refers to establishing a functionality in a running state. HTNs naturally enable encoding knowledge at different levels of abstraction; we can formulate tasks and encode high-level strategies in the methods of these tasks before reasoning on low-level primitive tasks~\cite{georgievski2015:htn}. 

We encode each abstraction as a compound task, namely \begin{footnotesize}\verb|install|\end{footnotesize} and \begin{footnotesize}\verb|run|\end{footnotesize}. Their methods encode specific configuration processes. One such method encodes the prerequisites for port activation. If the current component instance has require ports that are not active, the method first activates each port and makes a recursive call until all necessary ports are activated. The actual process of port activation is encoded in a separate task, which not only activates a required functionality, but also finds and installs (or runs) a component instance that provides that functionality. An instance with active require ports can then use the functionalities of other components with active provide ports. This is achieved by another method that encodes the port binding. For this process, the method depends directly on the binding actions. In addition to the methods for port activation and binding, there is a method for the case when all require ports are active and bound. To address the satisfaction of all require ports, we use a universal quantifier with implication in the method for both tasks, \begin{footnotesize}\verb|install|\end{footnotesize} and \begin{footnotesize}\verb|run|\end{footnotesize}. In the case of \begin{footnotesize}\verb|run|\end{footnotesize}, we have to deactivate the ports that will be no longer provided by the instance in the installed state. The process of port deactivation is similar to the process of port activation and it uses port unbinding. The process of port unbinding is more complex than the binding one, and requires checking for constraint violation. That is, we have to take care of active provide ports bound to active require ports. We use a separate task to encode the port unbinding. The \begin{footnotesize}\verb|unbindPorts|\end{footnotesize} task does nothing when the port is bound and needed for the next transition. When all necessary constraints are satisfied, it unbinds a specific port and recursively calls itself.

There are methods in \begin{footnotesize}\verb|install|\end{footnotesize} and \begin{footnotesize}\verb|run|\end{footnotesize} that deal with the case when there are no required functionalities for an instance. This means that we need a transition which can be handled by installing the component instance directly. In the case of running an instance, we invoke the port deactivation task to ensure a valid transition to the running state.

The modelling of the transitions from a running state to an installed state and further to an uninstalled state is analogous to the encoding of the tasks we described so far.

Finally, we encode instance duplication as a separate method. The method makes sure that the current component instance is in a specific state and it has at least one provide port bound. Consequently, a new component instance is created either in an installed state or in a running state, depending on the type of configuration.

\section{Final Remarks}
Our HTN planning domain model encodes realistic Cloud deployment problems. Using this domain, one can generate a problem file by specifying components and ports as objects, component states and ports as predicates, currently deployed instances as predicates, current states of deployed instances as predicates, bindings as predicates, and initialising the domain function to some value. Listing~\ref{lst:problem} shows an example of a problem file for the deployment of Wordpress, and Listing~\ref{lst:plan} shows its plan. Finally, HTN planning problems with varying difficulty can be generate automatically by manipulating the states and ports of components, as described in~\cite{georgievski2017:cloudapps}.

\begin{lstlisting}[
	float=t,
	caption={HTN problem file},
	label={lst:problem},
	language=PDDL]
(define (problem p) 
 (:domain deployment)
 (:objects
  wordpress mysql apache2 - component
  httpd mysql-in mysql-up - port
 )
 (:init
  (installed-require wordpress httpd)
  (running-require wordpress httpd)
  (running-require wordpress mysql-up)
  (installed-provide apache2 httpd)
  (installed-provide mysql mysql-in)
  (running-provide mysql mysql-up)
  (= (instance-number) 0)
 )
 (:htn
  :tasks (run wordpress)
  :ordering ()
  :constraints () 
 )
)
\end{lstlisting}

\begin{lstlisting}[
	float=t,
	caption={Example plan for the problem in Listing~\ref{lst:problem}},
	label={lst:plan},
	language=PDDL]
1.  (createInstance w0)
2.  (createInstance w1)
3.  (start a1)
4.  (bind httpd w0 a1)
5.  (start w0)
6.  (createInstance m2)
7.  (start m2)
8.  (run m2)
9.  (bind mysql-up w0 m2)
10. (run w0)
\end{lstlisting}

\section{Acknowledgments}
We thank Faris Nizamic, Alexander Lazovik and Marco Aiello for the discussions on earlier versions of the domain. We also thank Gregor Behnke for the valuable insights on the domain encoding and for transforming the domain to a suitable specification for the IPC 2020 on HTN planning.

\bibliographystyle{aaai}
\bibliography{lit}

\end{document}